\documentclass[prl,floatfix,showpacs,twocolumn]{revtex4}
\usepackage{amsmath,amssymb,graphicx}

\begin{document}
\title{The Parisi formula completed}

 \author{V.  Jani\v{s}}

\affiliation{Institute of Physics, Academy of Sciences of the Czech
Republic, Na Slovance 2, CZ-18221 Praha, Czech Republic }
\email{janis@fzu.cz}

\date{\today}

\begin{abstract}
The Parisi formula for the free energy of the Sherrington-Kirkpatrick
model is completed to a closed-form generating functional. We first
find an integral representation for a solution of the Parisi
differential equation and represent the free energy as a functional
of order parameters. Then we set stationarity equations for local
maxima of the free energy determining the order-parameter function on
interval $[0,1]$. Finally we show without resorting to the replica
trick that the solution of the stationarity equations leads to a marginally
stable thermodynamic state.
\end{abstract}
\maketitle

Interest of physicists in spin glasses has nor been  abating during
last few decades. In spite of a tremendous progress in understanding
of, in particular, the mean-field theory of spin glasses achieved in
recent years, answers to a number of questions about physical
properties of spin-glass models have not yet been found with ultimate
validity. Although presently the major theoretical effort
concentrates on clarifying the relevance of the mean-field
solution for finite-dimensional systems, there still remain
unresolved issues on the mean-field level.

The paragon mean-field theory for spin glasses is provided by the
Sherrington-Kirkpatrick (SK) model introduced more than thirty years
ago \cite{Sherrington75}. It took only a few years before Parisi
inferred by a specific ansatz the form of a consistent solution of
this model \cite{Parisi80b}. Parisi's solution, however, did not mark
the end but rather the beginning of interest of theorists in
spin-glass systems. There were two major reasons for further
investigations of properties of the SK model and the Parisi solution.
First, although most of the properties of the Parisi solution
indicated that it is the exact solution of the SK model, no
mathematical proof existed in that time. Second, the Parisi solution
was derived via the formal replica trick and the order parameters
necessary for its description were constructed from non-measurable
mathematical objects. The proper physical meaning of the
replica-symmetry breaking (RSB) in the spin-glass phase was initially
unclear.

The principal breakthrough in the proof of the exactness of the Parisi
solution was achieved only a few years ago by Guerra and Talagrand
\cite{Guerra03,Talagrand06}. They succeeded in proving rigorously
that the replica-symmetry scheme of Parisi represents simultaneously
both a lower as well as an upper bound on the free energy of the SK
model in the thermodynamic limit. The existence of the thermodynamic
limit with the self-averaging property of the free energy had already
been proved earlier \cite{Guerra02}. Unfortunately, neither the Parisi
formula for the free energy nor the construction of Guerra and
Talagrand do provide equations determining uniquely the thermodynamic
state in the spin-glass phase. Although Talagrand conjectures the
existence of a unique macroscopic state \cite{Talagrand03}, the
Parisi free energy is expressed only in a loose form of a formal
maximum in a large functional  space of order-parameter functions.
The way how to  approach practically the maximum and how does the
order-parameter function look like remain presently
unset. Moreover, not having an explicit representation for the
free energy stationary with respect to (w.r.t.) order parameters, it
is not straightforward to define physical quantities such as magnetic
susceptibility, entropy or specific heat.

The aim of this Letter is to extend the Parisi expression for the free
energy of the SK model to an explicit  closed-form functional of the order
parameters. Maximizing the free energy, unspecified in
the Parisi formulation, will then become a uniquely defined process of
finding solutions to stationarity equations determining fully the
actual form and values of all  order parameters. The Parisi free
energy extended in this way becomes a functional containing the
entire physical information and generating all  physical
quantities via (functional) derivatives without referring to the
replica trick and mathematical replicas. Moreover, our formulation of
the mean-field free energy opens a way to systematic expansions
and non-perturbative approximations to physical quantities in the
low-temperature spin-glass phase.
 
Using the replica trick, Parisi expressed the free energy of the SK
model as a functional of the order parameter function $q(x)$ for
$x\in [0,1]$ generalizing the single Sherrington-Kirkpatrick order
parameter $q =N^{-1}\sum_i m_i^2$, where $m_i$ are local
magnetizations. The free energy density in the Parisi solution is then
expressed as \cite{Parisi80b}
\begin{align}\label{eq:Parisi-FE}
F_T[q] &= - \frac {\beta^2}4 \left(1 + \int_0^1 q(x)^2 dx - 2
q(1)\right)\nonumber\\ &\qquad - \int_{-\infty}^{\infty}
\mathcal{D}\eta\ f(0,h + \eta \sqrt{q(0)})\ , \nonumber\\ F_P &=
\max_{q(x)}\ T F_T[q]
\end{align}
where we used  an abbreviation for a Gaussian differential
$\mathcal{D}\eta \equiv {\rm d}\eta\ e^{-\eta^2/2}/\sqrt{2\pi}$.  The
most difficult  "interacting" part of the above free energy $f(x,h)$
is not known explicitly. It is merely characterized by a Parisi
differential equation with an initial condition
\begin{align}\label{eq:Parisi-DE}
\frac{\partial f(x,h)}{\partial x} &= -\ \frac 12 \frac{dq}{dx}
\left[\frac{\partial^2 f(x,h)}{\partial h^2} + x \left(\frac{\partial f(x,h)}
{\partial h} \right)^2 \right]\ ,\nonumber \\
f(1,h) & = \ln\left[2\cosh \beta h\right]
\end{align}
the physical functional must obey. The physical solution for the free
energy should then be constructed by picking up the function $q(x)$,
being nondecreasing on interval $[0,1]$, so that functional
$f(x,h)$ obeying Eq.~\eqref{eq:Parisi-DE} maximizes the free energy
from Eq.~\eqref{eq:Parisi-FE}. We, however, do not know whether the
maximizing order-parameter function $q(x)$ obeys a specific equation
and if yes, how does the equation look like.
 
We can gain some insight into the phase space of the order-parameter
functions from the Guerra and Talagrand construction. It relies on
the so-called discrete replica-symmetry breaking scheme. The latter
can be derived straightforwardly from demanding thermodynamic
homogeneity of the resulting free energy. Thermodynamic homogeneity
is tested by stability of free energies with replicated spin systems
w.r.t. a weak interaction between the replicated spins
\cite{Janis05c}. We replicate the original system so many times
until we reach thermodynamic homogeneity, that is, independence of
a further replication. In this way a hierarchical structure of free
energy emerges due to successive replications of the original system.
The averaged free energy density with $K$ hierarchies can then be
represented as a functional of local response functions
to the inter-replica interaction \cite{Janis05c}
%\begin{widetext}
\begin{subequations}\label{eq:mf}
 \begin{multline} \label{eq:mf-avfe}
f^K(q,\Delta\chi_1,\ldots,\Delta \chi_K; m_1,\ldots,m_K) = - \frac
1\beta \ln 2 \\ + \frac \beta 4 \sum_{l=1}^K m_l\Delta\chi_l\left[2\left(q
+  \sum_{i=l+1}^{K}\Delta\chi_{i}\right) + \Delta\chi_l\right]\\
-\frac\beta 4 \left(1-q -\sum_{l=1}^K \Delta\chi_l\right)^2  - \frac 1\beta
\int_{-\infty}^{\infty} \mathcal{D}\eta\ \ln  Z_K\ .
\end{multline}
%\end{widetext}
We used a sequence of partition functions
\pagebreak[1]
\begin{equation}\label{eq:mf-hierarchy}
Z_l =  \left[\int_{-\infty}^{\infty}\mathcal{D}\lambda_l\
Z_{l-1}^{m_l}\right]^{1/m_l}\ ,
\end{equation}
\end{subequations}
the initial condition for which reads\\ $Z_0 = \cosh\left[\beta\left(h
    + \eta\sqrt{q} + \sum_{l=1}^{K}\lambda_l \sqrt{\Delta\chi_l}
  \right)\right]$. We again denoted the Gaussian differential
$\mathcal{D} \lambda$ introduced in Eq.~\eqref{eq:Parisi-FE}. The
order parameter $q$ is the only one directly connected with local
magnetizations. The other ones, $1>\Delta\chi_1 \ge \Delta\chi_2
\ge\ldots \Delta\chi_K \ge0$ and $1 \ge m_1 \ge m_2\ge \ldots m_k\ge
0$ were introduced due to an induced interaction between replicated
spins. All the order parameters are determined from stationarity
equations locally maximizing free energy \eqref{eq:mf}. The number of
hierarchies $K$ is not an order parameter characterizing a saddle
point of the free energy. It is fixed from stability
conditions, that is, it is a number of steps needed for achieving
thermodynamic homogeneity \cite{Janis05c}.
 
It was actually the discrete form of the replica-symmetry breaking
that was used by Guerra and Talagrand to prove its exactness for the
free energy of the SK model. They proved that free
energy~\eqref{eq:mf} becomes exact for the set of pairs
$\{m_1,\Delta\chi_1, m_2,\Delta\chi_2,\ldots,m_K,\Delta\chi_K\}$ for
which it is maximal. It is not specified whether the set is finite or
infinite, how the parameters should be  distributed on the underlying
interval $[0,1]$, or whether they obey specific (stationarity)
equations. The extremum may well become a supremum reached only at the
boundary of the multidimensional phase space.

It is important to realize that the discrete free energy,
Eq.~\eqref{eq:mf}, and the continuous one, Eqs.~\eqref{eq:Parisi-FE}
and~\eqref{eq:Parisi-DE},  are not identical. First, the
former has two sets of order parameters $m_l$ and $\Delta\chi_l$ for
$l=1,\ldots,K$ while the latter only one, $q(x)$ for $x\in[0,1]$.
Second, the order parameters from the discrete hierarchical free
energy are determined from stationarity equations unlike the Parisi
free energy, where the equation for $q(x)$ is unknown. Third, the
discrete free energy generally does not obey the Parisi differential
equation~\eqref{eq:Parisi-DE}. In fact, the Parisi free energy
emerges from a specific limit of the discrete ones, namely when
$K\to\infty$, $\Delta\chi_l = \Delta\chi/K \to dx$, and we neglect
second and higher powers of $\Delta\chi_l$ with the fixed index~$l$
\cite{Janis05c}.
 
Parisi's solution is defined on a subclass of measures considered by
Guerra and Talagrand on which we look for a maximum (supremum). It
seems that at least for the SK model, continuous of the Parisis
solution measures form a complete space and the Parisi free
energy determines the exact, marginally stable  solution. We demonstrated
it explicitly in the asymptotic region below the critical temperature of
the spin-glass phase in zero magnetic field \cite{Janis06b}. On the other
hand, there are models, such as the Potts spin glass \cite{Elderfield83},
where a discrete one-step RSB appears to be stable on a finite temperature
interval \cite{Gross85}.
 
We demonstrate in this Letter that independently of where the absolute
maximum of the RSB free energy may lie, we can always construct a
solution with a continuous distribution of differences
$\Delta\chi_l$ and a single order-parameter function $m(\lambda)$ on
the defining interval $[0,1]$ determined from an explicit equation
for a local maximum of the free energy.

To formulate the continuous free energy and to fix $0$ and $1$ as ends
of the underlying interval on which the order-parameter function is
defined we introduce two physical order parameters $q$ and $X$. The
former corresponds to $q(0)$ and the latter to $q(1) - q(0)$ in the
Parisi solution. We do not use the sequence $1\ge m_1,\ge \ldots
m_K\ge 0$ to set the interval on which the order-parameter
function is defined as in Eq.~\eqref{eq:Parisi-FE}. We find it more
convenient to reverse the choice and use $\Delta\chi_l = Xd\lambda$
as the fundamental infinitesimal differential. Neglecting all higher
than linear powers of $\Delta\chi_l$, unless accompanied by a
compensating summation over the labeling indices, free
energy~\eqref{eq:mf} reduces to
\begin{multline}\label{eq:FE-continuous}
f(q,X; m(\lambda)) = - \frac \beta4 (1 - q -X)^2 - \frac 1\beta \ln
2\\ + \frac {\beta X}2 \int_0^1d\lambda\ m(\lambda)\left[q + X(1 -
\lambda) \right] - \frac 1\beta \left\langle g(1, h + \eta
\sqrt{q})\right\rangle_\eta
\end{multline}
where $\langle X(\eta)\rangle_\eta = \int_{-\infty}^\infty
\mathcal{D}\eta X(\eta)$. The principal achievement of this Letter is
an explicit integral representation of the interacting free energy
$g(1,h)$. Dropping all terms of order $O(d\lambda^2)$ in the
discrete hierarchy of partition sums $Z_l$  in the continuous limit
$K\to \infty$ with $\Delta\chi_l = \Delta\chi/K = Xd \lambda$
we finally obtain
%
%\begin{subequations}\label{eq:g}
\begin{align}\label{eq:g0}
g(1,h) &= \Bbb E_0(X,h;1,0)\circ g_0(h) \nonumber \\     &\equiv \Bbb
T_\lambda \exp\left\{\frac X2 \int_0^1 d\lambda \left[\partial_{\bar{h}}^2
\right.\right. \nonumber \\ &\qquad \left.\left. + m(\lambda) g'(\lambda;h +
\bar{h})\partial_{\bar{h}} \right] \right\} g_0(h + \bar{h})\bigg|_{\bar{h}=0}\ ,
\end{align}
where we used prime to denote a derivative w.r.t. the magnetic field
$h$, $g'(\lambda, h)\equiv \partial_h g(\lambda,h)$.
To reach a closed form for the continuous free energy we introduced a
"time-ordering" operator $\Bbb T_\lambda$ ordering products of
$\lambda$-dependent non-commuting operators from left to right  in a
$\lambda$-decreasing succession. The time-ordered exponential is then
defined as
 \begin{multline*} \Bbb T_\lambda \exp\left\{\int_0^1 d\lambda
\widehat{O}(\lambda)\right\} \equiv 1 \\  + \sum_{n=1}^\infty \int_0^1
d\lambda_1 \int_0^{\lambda_1}d\lambda_2\ldots\int_0^{\lambda_{n-1}}\!\! d
\lambda_n \widehat{O}(\lambda_1)\ldots \widehat{O}(\lambda_n)\ .
\end{multline*} Time-ordering operators are a standard tool in
representing quantum many-body perturbation expansions. The initial
condition for the $\lambda$-evolution in Eq.~\eqref{eq:g0} is the local
free energy $g_0(h) = \ln \left[\cosh\beta h\right]$. Unlike the Parisi
construction we develop the solution on the defining interval from zero to
one.

It is straightforward to verify that the interacting part of the free
energy $g(\lambda,h)$ obeys a Parisi-like equation
\begin{align}\label{eq:g0-DE}
\frac{\partial g(\lambda,h)}{\partial \lambda} &=  \frac {X}{2}
\left[\frac{\partial^2 g(\lambda,h)}{\partial h^2} + m(\lambda)
\left(\frac{\partial g(\lambda,h)} {\partial h} \right)^2 \right]\ .
\end{align}
The opposite overall sign of the r.h.s. of this equation compared
with Eq.~\eqref{eq:Parisi-DE} is connected with the reverted
evolution of the initial condition used here.
 
Evolution operator $\Bbb E_0(X;\mu,\nu)$ contains only polynomials in
powers of derivatives w.r.t. an auxiliary magnetic field replacing the
Gaussian integration over auxiliary random fields $\lambda_l$ in
Eq.~\eqref{eq:mf}. It is, however, a nonlinear operator that is why
we must  find analogous integral representations for functions
$g'(\lambda,h)$ and $g''(\lambda,h)$ appearing in
Eq.~\eqref{eq:g0-DE}.

From the definition of the evolution operator $\Bbb E_0$  we obtain
\begin{multline}\label{eq:E-equation}
\frac {\partial g(\lambda,h)}{\partial h} = \Bbb E_0(X,h;\lambda,0)\circ
g_0^\prime(h) \\ + \frac X2 \int_0^\lambda d\nu\ m(\nu)\Bbb
E_0(X,h;\lambda,\nu)\circ \left[g^{\prime} (\nu, h)\partial_{h} g^{\prime}
(\nu, h)\right]\ .
\end{multline}
A solution to this integral equation can again be represented via an
evolution operator and the $\Bbb T$-ordered exponential
\begin{align}\label{eq:g1}
g'(\nu,h) &= \Bbb E(X,h;\nu,0)\circ g_0^\prime(h) \nonumber \\
&\equiv \Bbb T_\lambda \exp\left\{ X \int_0^\nu d\lambda \left[\frac 12
\partial_{\bar{h}}^2 \right.\right. \nonumber \\ &\qquad \left.\left.
 + m(\lambda) g'(\lambda;h + \bar{h})\partial_{\bar{h}} \right] \right\}
  g_0'(h + \bar{h})\bigg|_{\bar{h}=0}\ .
\end{align}
%\end{subequations}
Analogously we derive an evolution operator for the second
derivative of the free energy
\begin{align}\label{eq:g2}
g''(\nu,h) &= \Bbb E_2(X,h;1,0)\circ g_0^{\prime\prime}(h) \nonumber
\\ &\equiv \Bbb T_\lambda \exp\left\{ X \int_0^\nu d\lambda \left[\frac
12
\partial_{\bar{h}}^2 \right.\right. \nonumber \\
&\qquad \left.\left. + m(\lambda)\partial_{\bar{h}}\ g'(\lambda;h +
\bar{h}) \right] \right\} g_0^{\prime\prime}
(h + \bar{h})\bigg|_{\bar{h}=0}\ .
\end{align}

Having an explicit representation for the free energy we can derive
stationarity equations for its local extrema. Free energy
$f(q,X;m(\lambda))$ is a function of static parameters $q$ and $X$
and a nonlinear functional of the dynamical order-parameter function
$m(\lambda)$. Vanishing of the free energy w.r.t.  infinitesimal
variations the static parameters leads to the following equations
\begin{subequations}\label{eq:stationarity}
\begin{align}\label{eq:stationarity-q}
q &= \frac 1{\beta^2} \left\langle g'(1,h_\eta)^2\right\rangle_\eta\
,\\ \label{eq:stationarity-X} X &= \frac 1{\beta^2} \left[
\left\langle \Bbb E(X,h_\eta;1,0)\circ g_0^\prime(h_\eta)^2
\right\rangle_\eta \right.\nonumber \\ &\qquad \left. - \left\langle
g'(1,h_\eta)^2\right\rangle_\eta\right]\ .
\end{align}
We denoted $h_\eta \equiv h + \eta\sqrt{q}$. Vanishing of the free
energy w.r.t. infinitesimal variations of function $m(\lambda)$ leads
to a functional equation
\begin{align}
\label{eq:stationarity-lambda}   \lambda &= \frac 1{\beta^2X} \left[
\left\langle \Bbb E(X,h_\eta;1,0)\circ
g_0^\prime(h_\eta)^2\right\rangle_\eta \right.\nonumber \\ &\qquad
\left. - \left\langle \Bbb E(X,h_\eta;1,\lambda)\circ
g'(\lambda,h_\eta)^2\right\rangle_\eta\right]\
\end{align}
\end{subequations}
valid for any $\lambda \in [0,1]$. Notice that
Eq.~\eqref{eq:stationarity-lambda} for $\lambda =0$ is trivial and
for $\lambda=1$ coincides with Eq.~\eqref{eq:stationarity-X}. Hence,
only equations for $0<\lambda<1$ serve for the determination of $m$ as
a function of the evolution parameter $\lambda$.

Free energy~\eqref{eq:FE-continuous} complemented with stationarity
equations~\eqref{eq:stationarity} defines a thermodynamic state of
the SK model for all input parameters. It is thermodynamically
consistent so that physical values of all internal parameters
specifying the thermodynamic state are determined self-consistently
from stationarity equations and the standard thermodynamic relations
hold. For instance, magnetic susceptibility reads
\begin{equation}\label{eq:susceptibility}
\chi_T = \frac 1\beta \left \langle  g''(1,h + \eta\sqrt{q})
\right\rangle_\eta \ .
\end{equation}
We do not have an integral representation such as in the Parisi
formulation \cite{Parisi80b}, since we do not use $q(x)$ as the
order-parameter function but rather $m(\lambda)$. Nevertheless, we
have an alternative implicit representation of $g''(\lambda,h)$ in
Eq.~\eqref{eq:g2}. It is, however, important that we do not need to
resort to the replica trick to define and calculate physical
quantities in the thermodynamic state described by free
energy~\eqref{eq:FE-continuous}.

One of the attractive features of the presented extension of the
Parisi free energy is a possibility to verify stability of the
solution of equations~\eqref{eq:stationarity}. To
this purpose we utilize the fact that
Eq.~\eqref{eq:stationarity-lambda} holds for all $\lambda\in [0,1]$.
Then also a total derivative of both sides w.r.t. $\lambda$ must be
equal. Employing properties of the evolution operator $\Bbb E$ we
find
\begin{multline}\label{eq:stability-derivative}
\frac d{d\lambda} \Bbb E(X,h;1,\lambda)\circ g'(\lambda,h)^2 \\ = - X
\Bbb E(X,h;1,\lambda)\circ g''(\lambda,h)^2\ .
\end{multline}
Using this result in Eq.~\eqref{eq:stationarity-lambda} we obtain a
generalization of the de Almeida-Thouless condition
\cite{Almeida78} for marginal thermodynamic stability
\begin{align}\label{eq:stability-averaged}
\beta^2 &= \left \langle \Bbb E(X,h_\eta;1,\lambda)\circ
g''(\lambda,h_\eta)^2 \right\rangle_\eta \ .
\end{align}
It is actually a functional equation for $\lambda\in [0,1]$. With some
effort we can verify that Eq.~\eqref{eq:stability-averaged} is a
continuous limit of stability conditions derived in
\cite{Janis05c,Janis06a}. We see that the Parisi solution of the SK
model satisfying Eqs.~\eqref{eq:stationarity} is marginally stable
with no negative eigenvalues of the nonlocal susceptibility or the
spin-glass susceptibility.

Free energy~\eqref{eq:FE-continuous}  with the interacting part from
Eq.~\eqref{eq:g0} is defined implicitly, since the evolution
operator $\Bbb E_0$ contains the solution $g'(\lambda,h)$. This
cannot be avoided, since the Parisi differential
equation~\eqref{eq:g0-DE} is nonlinear. We can, nevertheless use the
implicit integral representation of the evolution operator  to
an approximate computation of the free energy and other physical
quantities. The most straightforward way is to expand the $\Bbb
T$-ordered exponential in powers of the exponent. This practically
corresponds to a power-series expansion of the order-parameter
function $m(\lambda)$. Such an approach becomes asymptotically exact
near the de Almeida-Thouless instability line. We hence can analyze
the critical behavior without the necessity to come over to truncated
a model. If one finds an effective way how to systematically
generates higher-order terms of such a power expansion, a rather
accurate approximation in the entire spin-glass phase can be
accomplished. Another method for resolving the evolution
operator is to approximate the order-parameter function with
piece-wise constant functions. In this way we approximate the
continuous scheme by a discrete one, resembling the discrete RSB from
Eq.~\eqref{eq:mf}.

To conclude, we completed the Parisi formula for the free energy of
the SK model so that it acquires the standard form demanded by
the fundamental principles of statistical mechanics. The derived free
energy is a function of two numerical order parameters $q$ and $X$ and
a functional of an order-parameter function $m(\lambda)$ defined on
interval $[0,1]$. The physical values of these order parameters are
determined from stationarity equations for local extrema
(maxima) of the free energy. The free energy thus becomes a well
defined generating functional from which all physical quantities can
be derived via standard thermodynamic methods. There is no need to
resort to the replica trick and a representation via mathematical
replicas to identify measurable quantities. The
integral representation of the solution of the Parisi differential equation
demonstrates that the thermodynamic state of the SK model
 is marginally stable  and allows for  explicit
systematic and non-perturbative approximations of the
thermodynamics of mean-field spin-glass models.

\begin{acknowledgments}
  Research on this problem was carried out within a project
  AVOZ10100520 of the Academy of Sciences of the Czech Republic.
\end{acknowledgments}


\begin{thebibliography}{99}

\bibitem{Sherrington75} D. Sherrington and S.  Kirkpatrick, \newblock
  Phys. Rev. Lett. {\bf 35}, 1972  (1975).

\bibitem{Parisi80b} G. Parisi, \newblock J. Phys. A {\bf 13}, L115,
  1101 (1980).

\bibitem{Guerra03} F. Guerra,\newblock  Commun. Math. Phys.
\textbf{233}, 1 (2003).

\bibitem{Talagrand06} M. Talagrand, \newblock Ann. Math. \textbf{163},
221 (2006).

\bibitem{Guerra02} F. Guerra and F. L. Toninelli,\newblock  Commun.
Math. Phys. \textbf{230}, 71 (2002).

\bibitem{Talagrand03} M. Talagrand, \newblock C. R. Math. Acad. Sci.
Paris \textbf{337}, 625 (2003).

\bibitem{Janis05c} V. Jani\v s, \newblock Phys. Rev. B \textbf{71},
214403 (2005).

\bibitem{Janis06b} V. Jani\v s and A. Kl\'\i\v c, \newblock Phys. Rev.
B \textbf{74}, 054410 (2006).

\bibitem{Elderfield83} D. Elderfield and D. Sherrington, \newblock J.
  Phys.  C{\bf 16}, L497 (1983).

\bibitem{Gross85} D. J. Gross, I. Kanter, and H. Sompolinsky,
\newblock Phys. Rev. Lett. \textbf{55}, 304 (1985).

\bibitem{Almeida78} J. R. L. de Almeida and D. J. Thouless, J. Phys. A
\textbf{11}, 983 (1978).

\bibitem{Janis06a} V. Jani\v s, \newblock Phys. Rev. B \textbf{74},
054207 (2006).

%\bibitem{Parisi80a} G. Parisi, \newblock J. Phys. A {\bf 13}, 1887,
%  (1980).
\end{thebibliography}
\end{document}